\documentclass[sigconf,screen,nonacm]{acmart}
\usepackage{multicol}
\usepackage{multirow}
\usepackage{colortbl}
\usepackage{subcaption}

\usepackage[most]{tcolorbox}
\usepackage{dblfloatfix}
\usepackage{epstopdf}
\AppendGraphicsExtensions{.tif}

\usepackage{xspace}
\newcommand{\systemname}{\systemnamens\xspace}
\newcommand{\systemnamens}{StreetLens}

\AtBeginDocument{%
  \providecommand\BibTeX{{%
    \normalfont B\kern-0.5em{\scshape i\kern-0.25em b}\kern-0.8em\TeX}}}

\begin{document}

\title[\systemname]{\systemname: Enabling Human-Centered AI Agents for Neighborhood Assessment from Street View Imagery}
% Understanding Neighborhood Environments From Street View Imagery Using Vision-Language Models
%\systemname: Enabling Human-Centered Agentic Spatial AI for Neighborhood Assessment from Street View Imagery


\author{Jina Kim, Leeje Jang, Yao-Yi Chiang, Guanyu Wang, Michelle C. Pasco}
\email{[kim01479, jang0124, yaoyi, wan00523, mpasco]@umn.edu}
\affiliation{%
  \institution{University of Minnesota}
  \city{Minneapolis}
  \country{USA}
}

\renewcommand{\shortauthors}{Kim et al.}

\begin{abstract}
Traditionally, neighborhood studies have used interviews, surveys, and manual image annotation guided by detailed protocols to identify environmental characteristics, including physical disorder, decay, street safety, and sociocultural symbols, and to examine their impact on developmental and health outcomes. Although these methods yield rich insights, they are time-consuming and require intensive expert intervention. Recent technological advances, including vision language models (VLMs), have begun to automate parts of this process; however, existing efforts are often ad hoc and lack adaptability across research designs and geographic contexts. In this paper, we present \systemname, a user-configurable human-centered workflow that integrates relevant social science expertise into a VLM for scalable neighborhood environmental assessments. \systemname mimics the process of trained human coders by focusing the analysis on questions derived from established interview protocols, retrieving relevant street view imagery (SVI), and generating a wide spectrum of semantic annotations from objective features (e.g., the number of cars) to subjective perceptions (e.g., the sense of disorder in an image). By enabling researchers to define the VLM’s role through domain-informed prompting, \systemname places domain knowledge at the core of the analysis process. It also supports the integration of prior survey data to enhance robustness and expand the range of characteristics assessed in diverse settings. \systemname represents a shift toward flexible and agentic AI systems that work closely with researchers to accelerate and scale neighborhood studies. \systemname is publicly available at \href{https://knowledge-computing.github.io/projects/streetlens}{https://knowledge-computing.github.io/projects/streetlens}.
\end{abstract}



\maketitle

\begingroup
\renewcommand\thefootnote{}
\footnotetext{© Jina Kim et al. 2025. This is the author's version of the work. It is posted here for personal use. Not for redistribution. The definitive version was published in The 1st ACM SIGSPATIAL International Workshop on Human-Centered Geospatial Computing (GeoHCC '25), https://doi.org/10.1145/3764917.3771334.}
\addtocounter{footnote}{-1}
\endgroup

% "© [Owner] [Year]. This is the author's version of the work. It is posted here for your personal use. Not for redistribution. The definitive version was published in {Source Publication}, https://doi.org/10.1145/{number}."

\vspace{-.1in}
\section{Introduction}

\begin{figure*}[!b]
  \centering
  \includegraphics[width=0.99\linewidth]{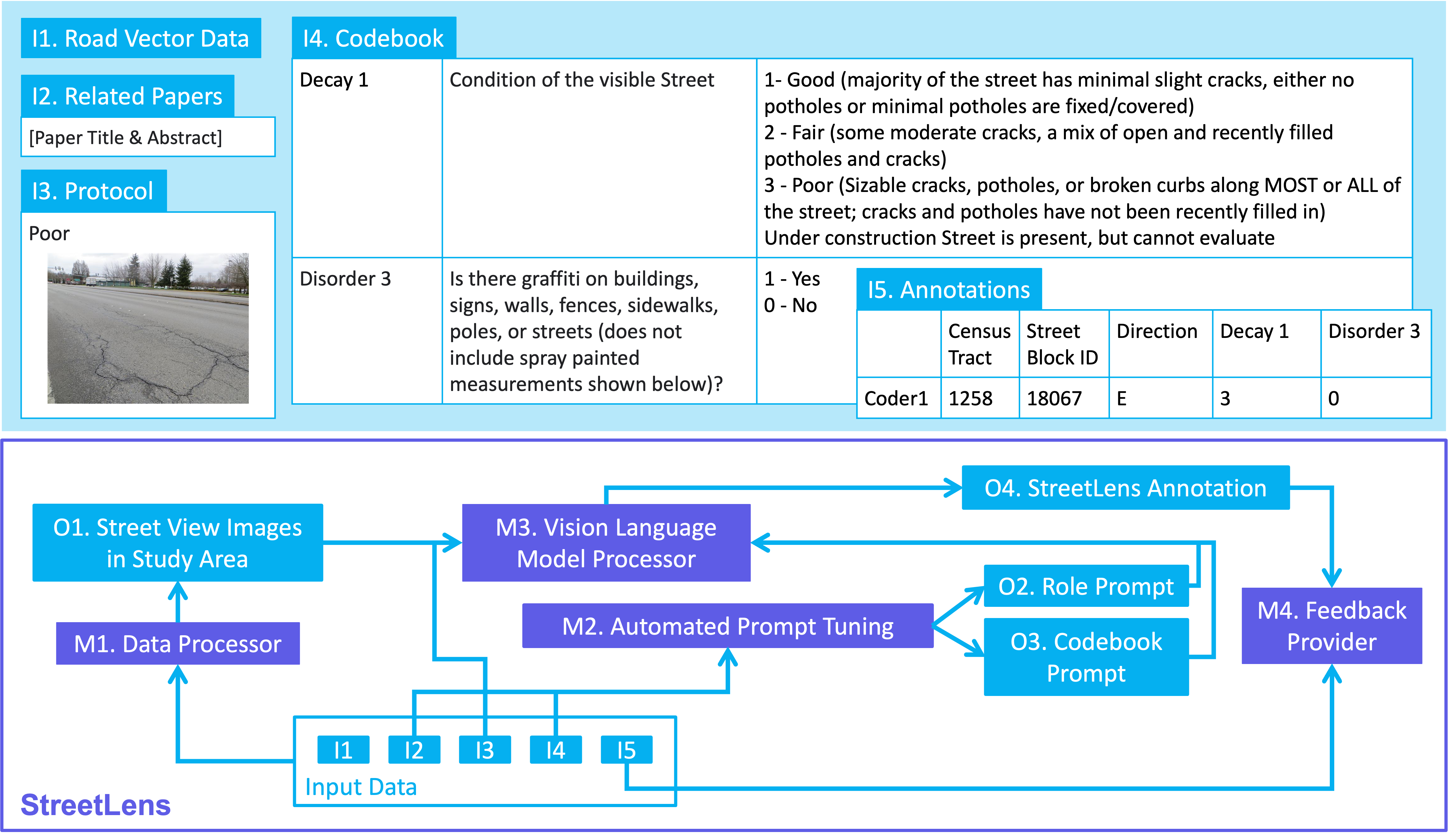}
  \vspace{-.1in}
  \caption{Example of input data for \systemname (top) and system workflow (bottom) illustrating \systemname architecture. \systemname comprises four modules (M\#). Input and output data are denoted as I\# and O\#, respectively. \vspace{-.1in}}\label{fig:workflow}
\end{figure*}

Neighborhood environments influence human well-being and have become a focus of research in urban studies~\cite{sampson1999systematic}, public health~\cite{bond2012exploring}, and family social science~\cite{pasco2024mixed}. Assessing neighborhood environmental characteristics is crucial to understand how neighborhood contexts influence outcomes such as adolescent development~\cite{pasco2024mixed}, mental health~\cite{liu2020natural}, and social cohesion~\cite{bjornstrom2014neighborhood, liu2020natural}. Neighborhood environment assessments have relied on qualitative and semi-quantitative methods, including surveys (e.g., Los Angeles Family and Neighborhood Survey)~\cite{bjornstrom2014neighborhood} and interviews~\cite{pasco2024mixed}. As street view imagery (SVI) becomes more accessible, researchers have created systematic social observation (SSO) protocols~\cite{odgers2012systematic, pasco2020mixed} to evaluate neighborhood environments using structured visual analysis using SVI. Protocols involve multiple human coders applying detailed codebooks to identify visual cues, such as signs of physical disorder, physical decay, and street safety. In practice, human coders rarely produce identical annotations for the same image, so multiple coders are needed per image, along with statistical analysis to compare their results. Despite offering invaluable insights, the process remains labor-intensive, dependent on trained experts, and challenging to scale across a wide variety of study contexts or geographic regions. 

Recent advances in vision language models (VLMs) have enabled researchers to automate certain aspects of the neighborhood environmental assessment process. Although researchers have begun to use VLMs with SVI~\cite{liang2024automatic, huang2024zero, jang2025multimodal}, these efforts often apply models to simple image annotation tasks, such as identifying objects in the physical environment, without a structured framework for adapting the models to fully assess the characteristics of a neighborhood. A limitation is the lack of a systematic method to ``teach'' VLMs in the way human coders are trained, including learning from the literature, coding protocols, and annotated examples. Also, VLM results are often accepted as is, without providing feedback to the researcher.

To address this gap, we introduce \systemname, an end-to-end researcher-centered workflow that mimics general human coder training processes. \systemname guides VLM through reviewing relevant studies, studying coding manuals, examining example annotations, and comparing the model output with those of experienced coders. \systemname automates the assessment pipeline using open-source VLMs and supports flexible adaptation to different study contexts and geographic regions. Specifically, \systemname allows researchers to incorporate domain knowledge as a central component of the workflow by leveraging relevant studies to configure the VLM through role prompting~\cite{schulhoff2024prompt}, explicitly defining the role of the VLM in the assessment process. \systemname then processes the codebook questions (for example, assessing physical disorder or identifying objects), retrieves the relevant SVI, and generates structured annotations aligned with expert-defined protocols. To support accessibility, we provide a Google Colab notebook that allows users to run \systemname with publicly available or user-provided image data, eliminating the need for advanced technical expertise. The result is a flexible and reusable workflow that facilitates assessing neighborhood environmental characteristics and benefits various studies across geographic settings.

\vspace{-.1in}
\section{Case Study}\label{sec:case_study}
% jina

This section introduces a case study from family social science that motivates the workflow of \systemname. The case study~\cite{pasco2020mixed} aims to assess the relationship between neighborhood environments and the use of ethnic and racial labels (i.e., defining social identity) by adolescents. To match ethnic and racial labels from semi-structured interviews with potential environmental features in neighborhoods,~\citet{pasco2020mixed} leverage the established SSO protocol~\cite{odgers2012systematic} (i.e., a detailed set of instructions) to train human coders. Multiple trained coders then use Google Earth Pro to virtually walk through each street segment, evaluating the degree of physical decay (e.g., deteriorated buildings, poor sidewalks) and sociocultural symbols (e.g., Spanish-language signs, Latino-owned businesses). 

After coding environmental features, \citet{pasco2020mixed} assess the reliability of the ratings by comparing coders' assessments of the street segments using intraclass correlation coefficients. This validation step removes outliers and ensures consistent ratings among coders, while minimizing personal bias. Involving multiple coders accounts for variations in individual perception and judgment, thus enhancing the overall quality of the assessments. Therefore, \systemname aims to serve as an additional agentic coder by enabling researchers to create an automated coder with domain-specific materials.

\section{\systemname}  
% jina
\systemname provides a researcher-oriented workflow (Figure~\ref{fig:workflow}), targeting users whose research involves neighborhood environmental assessments. A researcher begins using \systemname through a simple and guided interface. The first module, \textbf{M1. Data Processor}, prompts the researcher to upload key materials such as codebooks, protocols, related publications, and example annotations, as well as to specify the study area to retrieve the SVI data. This module organizes and prepares all inputs for the next steps. Next, \textbf{M2. Automated Prompt Tuning} uses the collected domain knowledge to define the role of the VLM agent and generate protocol-aligned prompts that follow the researcher’s coding instructions. These prompts are passed to \textbf{M3. VLM Processor}, which analyzes street-level images and generates assessments of environmental characteristics. The researcher then reviews the results using previous coding created by human coders. Finally, \textbf{M4. Feedback Provider} allows the VLM agent to provide explanations on how the VLM agent interpreted the coding instructions. This feedback helps the researcher understand the reasoning of the agent.

\paragraph{\textbf{M1. Data Processor}}
\systemname starts by asking the researcher to choose a study area, such as a city or specific census tracts, where they want to assess environmental characteristics. Based on the selected area, the system returns a set of predefined point locations. These points are derived from U.S. Census TIGER road data, which have been sampled at 5-meter intervals to support efficient data retrieval. For this paper, \systemname uses Google Street View imagery to match the source used in the original case study (Section~\ref{sec:case_study}) with human annotations. Once the study area is set, the system prompts the researcher to indicate which materials are available, such as a codebook, protocol, sample annotations, or academic papers in the domain that the user wants \systemname to focus on.

\paragraph{\textbf{M2. Automated Prompt Tuning}}
\systemname generates a role by reviewing related papers and assigns a role to the VLM to help it ``think'' like a trained human coder. This step enhances the VLM's performance~\cite{schulhoff2024prompt} and aligns with how human coders learn by reading background studies and understanding what to look for in the environment. To achieve this, \systemname uses the following prompt with the abstracts of relevant papers to the large language model (LLM):
\begin{quote}
\textit{You are an expert in the following fields and the author of the paper abstracts provided here: [I2. Abstracts of related papers]. Based on the expertise demonstrated, generate a general professional role description of yourself in one to two sentences, starting with "You are" written in the second person. This will be used as a system prompt introduction.}
\end{quote}

% \begin{quote}
% \textit{You are an expert in conducting mixed-methods research in urban sociology and ethnic studies, focusing on the impact of neighborhood environments on social behaviors and identity formation among Latinx adolescents. You specialize in using systematic social observations and qualitative interviews to compare and contrast the perspectives of researchers and adolescents, providing critical insights into the shared and unique aspects of urban environments and ethnic-racial identity within ethnically/racially segregated neighborhoods.}
% \end{quote}

% Leeje
After defining the specific role for the VLM agent, \systemname processes the codebook collected in M1, which contains questions and answer options for each environmental feature. \systemname first determines whether each pair of questions and answers presents a subjective or objective task. This step allows \systemname to align the VLM agent with the relevant task category. For example, in this case study (Section~\ref{sec:case_study}), \systemname distinguishes whether a given question and answer pair aims to assess subjective neighborhood perceptions or to verify the presence of specific objects (e.g., graffiti or bike lanes). \systemname uses the following prompt for this procedure:
\begin{quote}
\textit{%
You are a classifier of annotation tasks. 
Given a question and its answer options, decide if the task is perception (holistic/qualitative scene judgment such as condition/quality/intensity ratings) or object\_detection (presence, counting, or localization of specific object instances). 
Rules: If it asks to rate/assess overall condition or quality (e.g., Good/Fair/Poor), label as perception. 
If it asks to detect, count, or verify specific objects (e.g., cars, signs, pedestrians), label as object\_detection.
[I4. Question and answer options from codebook] Return only a single integer: 0 if perception, 1 if object\_detection. 
Do not include any words, JSON, spaces, or punctuation.
}%
\end{quote}

Then, for each question and its corresponding answer options, \systemname applies the following prompt to generate a codebook prompt:
\begin{quote}
\textit{
    Instruction: Rewrite the question as a clear, self-contained sentence, prefixed with "Question:". Then, rewrite each answer option as a full sentence explaining the meaning, starting with its number. Keep all numbers and meaning intact. Output plain text only, one sentence per line. [I4. Question and answer options from codebook] 
}
\end{quote}

\begin{figure}[h]
  \centering  \includegraphics[width=0.8\linewidth]{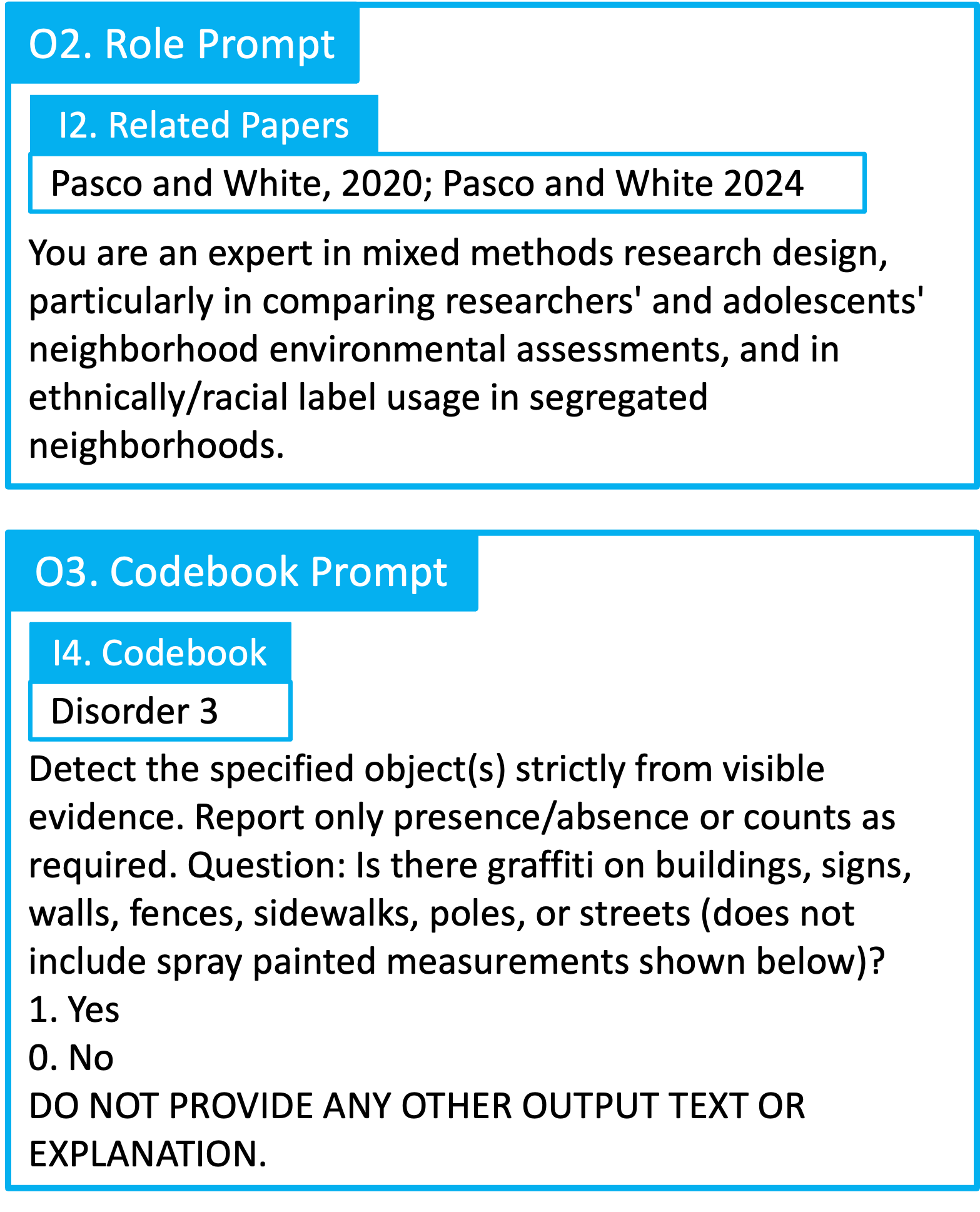}
  \vspace{-.1in}
  \caption{\systemname's automated prompt-tuning output (a domain-specific prompt) using related papers and the codebook.\vspace{-.1in}}\label{fig:prompt}
\end{figure}

For example, Figure~\ref{fig:prompt} shows the generated O2. Role Prompt and O3. Codebook Prompt. In this case, the LLM takes as input two studies on neighborhood environments and Mexican-origin adolescents~\cite{pasco2020mixed, pasco2024mixed}, along with the `Disorder 3' question and its answer options from the codebook.

\vspace{-.1in}

%, which guides the VLM to concentrate its evaluation within the appropriate task category. 
% The following is an example of a generated prompt for assessing the condition of a sidewalk (corresponding to the ``Decay 2'' code theme in the SSO protocol~\cite{pasco2020mixed}):
% \begin{quote}
% \textit{[Assigned role prompt] Assess the environmental condition of the sidewalk in the street view image provided, focusing on its surface and overall maintenance. Evaluate the condition of the sidewalk based on visual cues from the image and choose the best matching option. Options are: 1 - Good (NO holes, sizable cracks, or crumbling or uneven pavement) 2 - Fair (Holes, sizable cracks, or crumbling or uneven pavement, outgrown weeds along SOME of the side walk) 3 - Poor (Holes, sizable cracks, or crumbling or uneven pavement, outgrown weeds along most or ALL of the sidewalk) 99 - Under construction or cannot Evaluate. Your response must be an integer. DO NOT PROVIDE ANY OTHER OUTPUT TEXT OR EXPLANATION.}
% \end{quote}

% Figure~\ref{fig:prompt} shows an example prompt generated by \systemname based on input from the codebook, protocol, and relevant academic papers.

\begin{figure*}[h]
\vspace{-.1in}
  \centering
  \includegraphics[width=\linewidth]{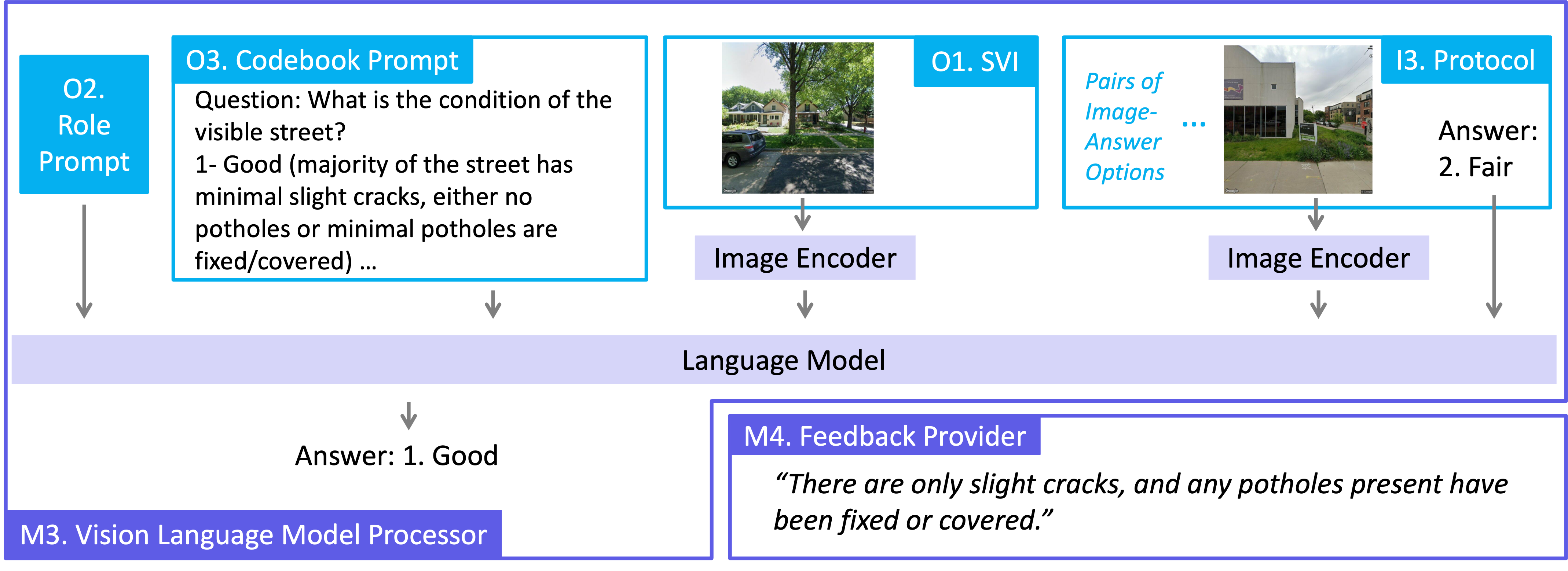}
  \vspace{-.1in}
  \caption{\systemname’s M3. Vision Language Model Processor and M4. Feedback Provider, showing both the input and the processed output. Refer Figure~\ref{fig:prompt} for O2. Role Prompt.\vspace{-.1in}}\label{fig:module3}
\end{figure*}

% As the prompt includes SVI as well, \systemname generates the <image> tags matching the number of processed SVI per prompt.

\paragraph{\textbf{M3. Vision Language Model Processor}} 

With the M2-generated prompts and the SVIs collected from M1, \systemname generates assessments for each environmental feature. The current version of \systemname loads the open-source lightweight VLM \textit{InternVL3-2B}~\cite{zhu2025internvl3}, which consists of an image encoder \textit{InternViT-300M-448px-V2\_5} with a language model \textit{Qwen2.5-1.5B}. Figure~\ref{fig:module3} presents how \systemname encodes a set of street view images using the image encoder. After the projection layer, \systemname passes the resulting image token embeddings, together with the tokenized text inputs from M2 prompts (i.e., O2 and O3), to the language model. In addition to the prompts generated in M2, \systemname leverages image–answer pairs from the I3 protocol, used as training materials for human coders, to perform in-context learning.

\paragraph{\textbf{M4. Feedback Provider}} After the VLM agent evaluates the environmental characteristics, \systemname provides feedback to researchers that includes explanations of the agent's evaluations, utilizing reasoning capabilities. For example, when coding the `Decay 1' measure (see Figure~\ref{fig:workflow} for measure details), which assesses the general condition of visible streets, \systemname provides the explanation of a certain street block ID \#281 assessment, `There are only slight cracks, and any potholes present have been fixed or covered' (Figure~\ref{fig:module3}). This process improves the interpretability of \systemname.

For demonstration, \systemname is available at \href{https://knowledge-computing.github.io/projects/streetlens}{https://knowledge-computing.github.io/projects/streetlens}, integrating data from the original case study (Section~\ref{sec:case_study}) and connecting to Cloudflare through a server located at the University of Minnesota. Users can individually toggle each module using buttons to explore StreetLens.

\section{Related Work}
Assessing neighborhood environment characteristics has traditionally involved trained coders conducting SSO and in-person audits. For example, a seminal study by~\citet{sampson1999systematic} employed video-based ratings of more than 23,000 street segments in Chicago to quantify physical disorder. To scale such efforts, researchers later adopted ``virtual audits'' using platforms like Google Earth and Street View~\cite{clarke2010using,Rundle2011StreetViewAudit,Odgers2012VirtualSSO}. These methods enabled remote assessment of neighborhoods with high inter-rater reliability and reduced cost. Based on this, computer vision techniques have been used to detect and quantify physical characteristics such as urban greenery and sidewalk quality~\cite{biljecki2021street} and to infer conceptual attributes of a higher level, such as perceived safety from images~\cite{naik2014streetscore, biljecki2021street}. However, such supervised approaches are often constrained by task-specific labels and limited generalizability to new contexts. Recent advances in VLMs offer more flexible solutions by enabling open-ended scene interpretation through joint image-text representations. VLM-based methods have been used to assess walkability~\cite{Blecic2024WalkabilityLLM}, generate structured descriptions of urban environments~\cite{Perez2025SAGAI}, and detect objects within audit categories (e.g., tree, vehicle, trash bin)~\cite{jang2025multimodal}, demonstrating promise in automating scalable, semantically rich neighborhood audits across diverse settings. In contrast to prior work tailored to specific studies, \systemname is a VLM-based, end-to-end, researcher-centered SSO workflow that simulates the human coder training process.
% , aiming to enhance adaptability across diverse study designs and geographic contexts.

% \vspace{-.2in}
\section{Discussion and Future Work}
We introduce \systemname, a human-centered workflow designed to enhance the assessment of environmental characteristics by acting as an additional coder. \systemname directs the VLM agent to replicate the training and evaluation process used by human coders in collaboration with domain experts, adhering closely to established assessment protocols. 
% The current implementation includes feedback that provides explanations of the VLM agent’s coding decisions.

In future work, we aim to make the workflow more human-centered to better assist researchers throughout the process. 
% This includes adding features that track the origin and history of data and annotations (provenance tracking), which helps researchers understand the source of their results and how they were generated. 
This includes adding a feedback loop that allows researchers to interpret \systemname's decisions, refine the assessment process, and explore different types of automated coders to determine which coders most closely mimic human coding. To evaluate \systemname, we plan to use intra-class correlation (ICC), treating the automated coder as one of the human annotators. ICC will provide feedback that lets researchers monitor whether \systemname's outputs remain reasonable and reliable, similar to the evaluation of new human coders. These enhancements will help researchers trust the automated assessments more and make it easier to review and refine the results.

\section*{Ethical Considerations}
We acknowledge that any machine learning models, including vision language models, can reproduce social biases, especially when interpreting sociocultural contexts in diverse neighborhoods. In future work, we will assess potential sources of bias and apply participatory design methods in collaboration with domain experts to ensure that \systemname remains a responsible and human-centered tool.

\bibliographystyle{ACM-Reference-Format}
\bibliography{biblio}

% \appendix

\end{document}